# Nonlinear Modeling Approach to Human Promoter Sequences

Huijie Yang[*,1], Fangcui Zhao[2], Jianzhong Gu[3], Binghong Wang[1]

[1] Nonlinear Science Center and Department of Modern Physics, University of Science and Technology of China, Hefei Anhui 230026, China

[2] Institute of life science & Bioengineering. Beijing University of Technology, Beijing 100022, China

[3] China Institute of Atomic Energy, P.O.Box 275(18), Beijing 102413, China

## ABSTRACT

By means of the nonlinear modeling technique, we find the nonlinear deterministic structures in the totally 4737 Human promoter sequences. These deterministic structures prefer to occur much more outside of a special region around the transcriptional start site. The number and positions of the deterministic structures are basically different for different promoter sequences. Generally, they do not coincide with the CpG islands, the simple repetitive sequences and the low-complexity sequences, which tell us that they should be special structures with biological functions rather than these special segments.

[*] To whom correspondence should be addressed. E-mail: huijieyangn@eyou.com

## I. INTRODUCTION

Understanding gene regulation is one of the most exciting topics in molecular genetics [1]. To learn how the interplay among thousands of genes leads to the existence of a complex organism is one of the great challenges. The quantity of information gained in the sequencing and gene expression projects both requires and enables us to use computers to solve this problem [2-4]. Promoter sequences are crucial in gene regulation. The general recognition of promoters and the analysis of these regions to identify the regulatory elements in them are the first step towards complex models of regulatory networks.

Experimental knowledge of the precise 5' ends of cDNAs should facilitate the identification and characterization of regulatory sequence elements in proximal promoters [5]. As a first step in this direction, Suzuki et al. used the oligocapping method to identify the transcriptional start sites from cDNA libraries enriched in full-length cDNA sequences, which they have made available at the Database of Transcriptional Start Sites (http://dbtss.hgc.jp/) [6]. Consequently, the authors in reference [7] have used this data set and aligned the full-length cDNAs to the human genome, thereby extracting putative promoter regions (PPRs). Using the known transcriptional start sites from over 5700 different human full-length cDNAs, a set of 4737 distinct PPRs are extracted from the human genome. Each PPR consists nucleotides from -2000 to +1000 bp, relative to the corresponding transcriptional start site. They counted eight-letter words within the PPRs, using z-scores and other related statistics to evaluate the over- and under- representations.

In general, the promoter is understood as a combination of different regions with different functions [8-11]. The sub portion of the promoter surrounding the transcription start site, called core promoter, interacts with RNA polymerase II and basal transcription factors. It is the minimal sequence required for initiating transcription. Few hundred base pairs upstream of the core promoter are the gene-specific regulatory elements (proximal promoter region), which are recognized by transcription factors to determine the efficiency and specificity of promoter activity. Far distant from the transcription start site there are enhancers and distal promoter elements which can considerably affect the rate of transcription. Detailed researches have shown that multiple binding sites contribute to the functioning of a promoter, with their position and context of occurrence playing an important role. Several classes of interspersed repetitive DNA derived from mobile genetic elements account for at least 45% of the human genome [12,13] and 39% of the mouse genome [14]. Recent reports indicated that in these so-called "junk DNA" specific elements have evolved a biological function by donating transcriptional regulatory signals, including alternative promoters, to cellular genes [15-17]. Large-scale studies determine that repeats do indeed participate in the regulation of numerous human and mouse genes [18]. It is found that nearly 25% of regulatory regions in the Human Promoter Database (http://zlab.bu.edu/~mfrith/HPD.html) contain transposable element sequences [19]. Combined, these studies suggest that the promoter's biological function is a cooperative process of different regions such as the core promoter, the proximal promoter region, the enhancers/silencers, the insulators, the CpG islands and so forth. But how they cooperate with each other is still a problem to be investigated carefully.

The correlations in DNA sequences have been an important topic in bioinformatics [20]. It is believed that the correlations in bio-sequences may correspond with biological functions. Peng et al. [21] and Li and Kaneko et al. [22] reported first the long-range correlations in DNA sequences. Using the DNA walk method and the $1/f$ spectrum, it is demonstrated that non-coding sequences carry long-range correlations, while coding segments are compatible with short-range correlations. These novel works stimulated many following similar findings [23]. Very recently, along these lines an investigation was conducted on the inter-distance distribution between consequent occurrences of identical oligonucleotides (5 or 6-bases nucleotide). It is found that the inter-distance distributions for the special oligonucleotides containing consensus sequences of an RNA polymerase II promoter are compatible with the existence of long tailed distributions, while that for the random oligonucleotides follow short-tailed distributions. An earlier accordant observation has also been presented in literature [24,25] that the size distribution of coding sequences and non-coding sequences follows short-range distributions as Gaussian or exponential and long-range distributions as power-law, respectively. By means of the theoretical models of DNA, the dynamical activities are simulated [26-29], which show us that the part of the promoters where the RNA transcription has started are more active than a random portion of the DNA. All the evidences suggest that the segments between the transcription regularity-related oligonucleotides in the human PPR sequences should have special biological functions on the cooperation process of the core promoter, the proximal promoter, the enhancers and so on. The statistical properties of the human PPR sequences, especially the correlation characteristics, may shed light on this problem.

In two stimulating papers [30,31], the nonlinear modeling technique (NM) is used to find the deterministic structures in non-coding regions and in Alu repeats. The NM method is initially designed to distinguish between chaos and noise in time series [32-35]. The key idea is to explore quantitatively similarity along the sequence at sub-chain vicinities of equal sub-chains of size (the embedding dimension). It can detect the deterministic structures, i.e., the predictability in a series, which cannot be detected by the standard methods, such as Fourier transformation and power spectrum. Reasonable information can be obtained even for very short sequences, to cite an example, distinguishable result can be obtained for Alu repeats with 100~200 base pairs from HUMHBB DNA sequences [31]. It is found that the non-coding sequences have deterministic structures, especially in human DNA sequences, while the coding sequences behave similar with randomly positioned sequences. But the discussions are based upon an average over the whole length of the considered sequences. The local properties are not considered at all.

In this paper we use the NM method to reveal information embedded in the set of 4737 PPRs extracted from the human genome presented in the Web site in Ref. [7]), based upon which we find the sequence segments with deterministic structures. Rather than the global average properties, we present in this work the local characteristics of the PPRs. Instead of the occurrence probabilities of the special oligonucleotides containing consensus sequences of promoters, we are interested in the deterministic structures in the whole sequences.

## II. METHOD

To make our paper self-contained, we review NM method briefly. A schematic representation of the method is shown in Fig.1. Consider a PPR denoted with $(x_1, x_2, x_3, \cdots x_N)$. Each data $x_k$ will consist one of the four symbols A, C, G or T. All the possible sub-chains with length $M$ read,

$$X_k = (x_k, x_{k+1}, \cdots x_{k+M-1}) | k = 1, 2, 3, \cdots, N - M + 1. \tag{1}$$

The properties of each sub-chain can be employed to represent the corresponding local characteristics.

For each sub-chain $X_k$, connecting the starting and the end of this segment, we can organize it in $d$-dimensional delay-register vectors:

$$X_k^1 \equiv (x_k, x_{k+1}, \cdots x_{k+d-1})$$

$$X_k^2 \equiv (x_{k+1}, x_{k+2}, \cdots x_{(k+1)+d-1})$$

$$\vdots$$

$$X_k^M \equiv (x_{(k+M-1)}, x_{(k+M-1)+1}, \cdots x_{(k+M-1)+d-1}) \tag{2}$$

For each vector $X_k^i \equiv (x_{k+i-1}, x_{k+i}, \cdots x_{(k+i-1)+d-1})$, one searches for its nearest neighbor $X_k^{H(i)}$ and then compares how close the symbols $x_{[(k+i-1)+d-1]+1}$ and $x_{[(k+H(i)-1)+d-1]+1}$ are following these two vectors.

The closeness of a pair of symbols is measured by the Hamming distance,

$$u(x_u, x_v) = \begin{cases} 0, & x_u = x_v \\ 1, & x_u \neq x_v \end{cases}, \tag{3}$$

The closeness of a pair of vectors $X_k^i$ and $X_k^j$ can then be measured by,

$$U(X_k^i, X_k^j) = \sum_{m=0}^{d-1} u(x_{k+i-1+m}, x_{k+j-1+m}). \tag{4}$$

The nearest neighbor $X_k^{H(i)}$ of a given vector $X_k^i$ is randomly selected among the solutions of $U(X_k^i, X_k^j)$ such that it is a minimum for $i \neq j$. The overall mean error in the sub-chain $k$ can be computed as,

$$E(M,k,d) = \frac{1}{M} \sum_{i=1}^{M} u(x_{[(k+i-1)+d-1]+1}, x_{[(k+H(i)-1)+d-1]+1}). \tag{5}$$

For uncorrelated chains, there is no relation between any pair of bases. The overall mean error in Eq. (5) can be approximated by,

$$E_{uncorrelated} = \rho(A)[1-\rho(A)] + \rho(T)[1-\rho(T)] + \rho(C)[1-\rho(C)] + \rho(G)[1-\rho(G)], \tag{6}$$

where $\rho(\cdot)$ denotes the probability of occurrences for the symbol $(\cdot)$. The finite length of a DNA segment will induce a deviation from this theoretical result. The criterion by which we find deterministic structures is then,

$$E(M,k,d) < E_{criterion}(M,k,d) \equiv E_{uncorrelated} - \Delta(M,k,d), \tag{7}$$

where $\Delta(M,k,d)$ is the interval between the theoretical value of $E_{uncorrelated}$ and the minimum of the possible values of $E_{uncorrelated}$ induced by finite length. Shuffling the considered segment for $J$ times ($J$ should be enough for statistical analysis), $\Delta(M, k, d)$ can be estimated as,

$$\Delta(M,k,d) = E_{uncorrelated} - \min\{E_t(M,k,d) | t = 1, 2, 3, \cdots J\}. \tag{8}$$

An alternative solution to distinguish the deterministic structures is presented in Ref. [31]. The ratio $\gamma = \dfrac{E(M,k,d)}{E_{uncorrelated}}$ is used and according to the calculations with fewer sample sequences the criterion is set to be $\gamma \leq \gamma_c = 0.85$. Empirical result shown in Table (1) tells us that the value of $\gamma$ varies in a considerable large region, and even can reach $78\%$ for a special set of the occurrence probabilities. Hence, the solution the present work may be relatively better due to the large set of samples ($J$ is assigned $3000$).

The other important feature in the NM method is the sensitivity to detect the deterministic structures. Consider a segment with deterministic structure as depicted in Fig.1(c), whose length is $L$. There is a delay interval that only when the overlap between the sliding window and the considered segment is larger than it can we find a distinguishable decrease of the overall average error. This interval is denoted as $s$. In the calculations, the region from $A$ to $B$ is used to represent the segment with deterministic structure. Clearly, we have,

$$\begin{cases} B - A = L + M - 2s, \\ L = c_l \cdot s > s, \\ M = c_m \cdot s > s. \end{cases} \tag{9}$$

Let $\eta = c_l + c_m - 2 > 0$, which leads,

$$B - A = \eta s = \dfrac{\eta}{c_l} L = \dfrac{\eta}{c_m} M. \tag{10}$$

Generally, we have $1 < c_m < 4$. Hence there is a linear relation between the real length of the deterministic structure and the corresponding representative length, $B - A$, in the calculations. The segments with length $L < \dfrac{M}{c_m}$ cannot be detected.

## III. RESULTS AND DISCUSSIONS

All the 4737 PPRs from the human genome are considered in this paper. Consider the total length of a PPR as a sub-chain, that is, $M = N$ and $k = 1$. For all the PPRs, we calculate the overall mean errors as a function of the embedding dimension $d$. In Fig.2 the results for the sequences labeled 1,10,100,1000,1001,1002,1004,1005,1006,1007,1008,1009 in the web site in Ref. [7] are presented as the typical examples, denoted with sequence 1 to sequence 12, respectively. While for uncorrelated sequences the overall error does not dependent on the dimension $d$, the result for a PPR decreases rapidly with the increase of $d$ from 1 to a critical value $d_c$, where the overall error reaches the minimum value. From $d_c$ the overall error will increase slowly. The existence of the critical value $d_c$ tells us that the considered sequences behave multiple predictabilities. In the scale of $d < d_c$, the sequences are much predictable compared with that in the scale of $d > d_c$. The overall error with a dimension near $d_c$ can be employed as the representation of the deterministic structures in a PPR sequence. The distribution of $d_c$ for all the PPRs is shown in Fig.3. We can find that the number of PPRs with $9 \leq d_c \leq 17$ is $3724$, covering $78.6\%$ of the total number of 4737. As a balance of the above features, in the following discussions the value of $d$ is assigned the average value of $d_c$, $d = \langle d_c \rangle = 15$, which is also consistent with the findings in Ref. [30].

The other adjustable parameter is the length of a sub-chain. If it is too long, the overall error will be an average of a large scale, which cannot give us local information precisely. While for a too short sub-chain, the local information will be submerged in large fluctuations. Results for all the PPRs are calculated and Fig.4 shows several typical results of the overall error $E(M, k, d)$ versus the position $k$. The results for the corresponding shuffling sequences are also presented. The local characteristics of a PPR are different completely. For the most parts the elements are arranged randomly and there are not deterministic structures. There are also several special parts where the overall errors are significantly smaller than that of the corresponding shuffling ones. These valleys are sensitive to the length of a sub-chain $M$. With the decrease of $M$, the bottoms of the valleys will become smaller and smaller and the borders become much more distinguishable.

There exist a suitable value of $M$, at which we can obtain the valleys in an acceptable precision and the shape is smooth enough. The randomly positioned parts are not sensitive to the length of a sub-chain $M$, where the overall errors fluctuate around that of the corresponding shuffled sequences.

A suitable value of a sub-chain $M$ is important for the other reason. As shown in the results of the PPRs numbered sequence3, sequence4 and sequence5, some valleys can be found only with larger value of $M$. $M = 500$ may be a proper choice in the present calculations.

Assigning $M = 500$, the overall errors as a function of the position $k$ for different values of the dimension $d$ are calculated for all the PPRs. Fig.5 shows several typical results. It is found that the valleys are sensitive to the change of $d$, while the uncorrelated segments are not sensitive to it at all.

Selecting $J = 3000$, we can obtain the criterion values for different sets of probabilities of the occurrences for the base pairs $A, T, C$ and $G$, as shown in Table (1). The value of $\frac{E_{criterion}(M,k,d)}{E_{random}}$ varies in a considerable large region (up to $78\%$ for a special set of the occurrence probabilities). This is the reason why we present this new solution instead of that suggested in reference [31].

Some of the valleys have deterministic structures, called deterministic valleys (DVs) in this paper. The DVs in all the PPRs are detected. Fig.6 presents a typical result to determine the position and length of a DV. The DVs can be identified by comparing the overall errors of the segments with the corresponding criterions. In the solid line the points whose values are 1 denote the segments having deterministic structures. A DV is a set of successive points in the solid line whose values are 1. In all the 4737 PPRs, the positions and the lengths of the DVs vary significantly. Fig.7 shows the probability distribution function of the bottom lengths of the DVs （denoted with $L_{bottom}$, it is also the representative length B-A in Eq.(10)）. With the increase of length this probability decreases rapidly in a power-law way, i.e., $P(L_{bottom}) \sim L_{bottom}^{-1.91}$. It is reasonable to believe that the DVs should have some special biological functions. The parameter values are $(M,d) = (500,15)$.

The occurrence probabilities of these DVs at different positions are estimated relative to the number of all the sequences, denoted with $p_{DV}$, as presented in Fig.8. To dismiss the large fluctuations we present the integrated probability $Q(k) = \sum_{m=1}^{k} p_{DV}(k)$ also. There is a special region near the transcriptional start site $T = 2001$, i.e., $[1500, 2000]$, where the integrated probability increases with a significantly slow speed. Consequently, the occurrence probabilities in this region should be small significantly comparing with that outside of this region. Filtering out the DVs with bottom lengths less than $10, 50$ and $100$, the integrated probabilities are also illustrated in Fig.8, respectively. We can find that the DVs with long bottom lengths prefer much more to occur outside of this special region.

One of the findings in Ref. [7] is that many transcription factor-binding sites in the human PPRs such as the TATA, GC and CAAT boxes often occur near the transcription start site [36]. They have preferred locations between the positions $1700$ and $2050$ bp (i.e., $-300$ and $+50$ relative to the transcriptional start site), which suggests that the basal promoter and nearby upstream regulatory elements are found in the region between $1700$ and $2050$ bp, in accord with a recent study from the Myers laboratory, where 91% of 152 DNA fragments containing regions $1450$ to $2050$ were active as promoters in at least one of four cell types evaluated [5]. Hence, the special region of $[1500, 2000]$ where the occurrence probability of DVs tends to vanish in our findings should correspond with the region that the transcription factor-binding sites cluster in. The transcription factor-binding sites are identified from the vertebrate matrices in the TRANSFAC® Professional Suite (Version 7.2) [37], using the TFBS Perl modules for TFBS analysis [38] (http://forkhead.cgb.ki.se/TFBS/).

From Eq.(9) we have $L_{bottom} = \dfrac{500 \cdot \eta}{c_m}$ and the sensitivity is $L > s = \dfrac{500}{c_m}$. The real lengths of the deterministic structures found in this paper should be at least several hundreds. Hence, it is necessary to distinguish the DVs from the other segments having special structures as the CpG islands, the simple repetitive sequences and the low-complexity sequences.

In Ref. [7], a DNA region is defined to be a CpG island, if (i) it is $\geq 500$ bp in length; (ii) it has a G + C content $\geq 50\%$; and (iii) its ratio of CpG observed/expected is greater than or equal to $0.6$. They find that in the total of 4737 PPRs the CpG islands appear to cluster near the transcriptional start site (the position $2001$). The occurrence probabilities of the DVs are different completely with that of CpG islands, which can tell us that the DVs found in the present work should be special structures rather than the CpG islands.

The simple repetitive and low-complexity sequences are absent from the proximal promoter regions, but can often be found in the upstream sequences. The program Repeatmasker Web Server [39] is used to search the simple repeats and the low-complexity segments in all the PPRs. Fig.9 shows the length distribution of the found segments and the occurrence probabilities at different positions. The length distribution obeys a power-law, i.e., $P(L_{repeat}) \propto L_{repeat}^{-2.35}$. The occurrence probability at different positions $p_{repeat}(k)$ distributes homogenously in all the positions except the special range $[1850,1980]$, where the proximal promoter regions prefer to occur and a significant sharp peak is shaped. Comparison with the DV distribution shows that the DVs are not the simple repetitive or low-complexity sequences.

In summary, in each PPR sequence we find some special regions where we can find significant deterministic structures, while outside of these regions the elements are positioned in an uncorrelated way. These deterministic structures are called deterministic valleys (DVs) in this paper. The number, positions and widths of the DVs vary for different PPRs. A suitable set of the parameters $(M,d)$ can reveal the DVs almost perfectly. The nontrivial occurrence probability of the widths and the positions of these DVs tell us that they may have new important biological functions rather than CpG islands, the repetitive sequences and the low-complexity DNA, which should be investigated in detail. The relations of the DVs with the other segments as the core promoters, the enhancers, the transposable elements (TE) and so forth are an interesting problem to be investigated in detail.

**ACKNOWLEDGEMENTS**

This work was supported by the National Science Foundation of China (NSFC) under Grant No.70571074, No.70471033. One of the authors (H. Yang) would like to thank Prof. Yizhong Zhuo in China Institute of Atomic Energy for Stimulating discussions.

**REFERENCES**

[1] Ohler,U. and Niemann,H. (2001) Identification and analysis of eukaryotic promoters: recent computational approaches. Trends Genet., **17**, 56-60.

[2] Strausberg,R.L., Feingold,E.A., Grouse,L.H., Derge,J.G., Klausner,R.D., Collins,F.S., Wagner, L., Shenmen,C.M., Schuler,G.D., Altschul,S.F. et al. (2002) Generation and initial analysis of more than 15,000 fulllength human and mouse cDNA sequences. Proc. Natl Acad. Sci. USA, **99**, 16899-16903.

[3] Carninci,P., Waki,K., Shiraki,T., Konno,H., Shibata,K., Itoh,M., Aizawa,K., Arakawa,T., Ishii, Y., Sasaki,D. et al. (2003) Targeting a complex transcriptome: the construction of the mouse full-length cDNA encyclopedia. Genome Res., **13**, 1273-1289.

[4] Lander,E.S., Linton,L.M., Birren,B., Nusbaum,C., Zody,M.C., Baldwin,J., Devon,K., Dewar, K., Doyle,M., FitzHugh,W. et al. (2001) Initial sequencing and analysis of the human genome. Nature, **409**, 860-921.

[5] Trinklein,N.D., Aldred,S.J., Saldanha,A.J. and Myers,R.M. (2003) Identification and functional analysis of human transcriptional promoters. Genome Res., **13**, 308-312.

[6] Suzuki,Y., Yamashita,R., Nakai,K. and Sugano,S. (2002) DBTSS:DataBase of human Transcriptional Start Sites and full-length cDNAs. Nucleic Acids Res., **30**, 328-331.

[7] Leonardo,M.–R., John,L.S., Gavin,C.K. and David,L. (2004) Statistical analysis of over-represented words in human promoter sequences. Nucleic Acids Research, **32**, 949-958. See also ftp://ftp.ncbi.nlm.nih.gov/pub/marino/published/hs_promoters/fasta/ .

[8] Werner,T. (1999) Models for prediction and recognition of eukaryoticpromoters. Mammalian Genome, **10**, 168-175.

[9] Pedersen,A.G.,  Baldi,P., Chauvin,Y., Brunak,S. (1999) The biology of eukaryotic promoter prediction - a review. Comput Chem , **23**, 191-207.

[10] Zhang,M.Q. (2002) Computational methods for promoter recognition. In: Jiang T, Xu Y, Zhang,M.Q., editors. Current topics in computational molecular biology. Cambridge, Massachusetts: MIT Press; p. 249-268.

[11] Narang,V., Sung,W.-K., Mittal A. (2005) Computational modeling of oligonucleotides positional densities for human promoter prediction. Art. Intel. Med., **35**, 107-119.

[12] Landry,J.-R., Mager,D.L. and Wilhelm B. T. (2003) Complex controls: the role of alternative

promoters in mammalian genomes. Trends. Genet. 19, 640-648.

[13] Lander, E.S. et al. (2001) Initial sequencing and analysis of the human genome. Nature 409, 860–921.

[14] Waterston, R.H. et al. (2002) Initial sequencing and comparative analysis of the mouse genome. Nature 420, 520–562.

[15] Landry, J-R. et al. (2002) The Opitz syndrome gene Mid1 is transcribed from a human endogenous retroviral promoter. Mol. Biol. Evol. 19, 1934–1942.

[16] Medstrand, P. et al. (2001) Long terminal repeats are used as alternative promoters for the endothelin B receptor and apolipoprotein C-I genes in humans. J. Biol. Chem. 276, 1896–1903.

[17] Landry, J-R. and Mager, D.L. (2003) Functional analysis of the endogenous retroviral promoter of the human endothelin B receptor gene. J. Virol. 77, 7459–7466.

[18] Rosenberg, N. and Jolicoeur, P. (1997) Retroviral pathogenesis. In Retroviruses (Coffin, J.M. et al., eds), pp. 475–586, Cold Spring Harbor Press.

[19] Jordan, I.K. et al. (2003) Origin of a substantial fraction of human regulatory sequences from transposable elements. Trends Genet. 19, 68–72.

[20] Buldyrev,S.V., Goldberger,A.L., Havlin,S., Mantegna,R.N., Matsa,M.E., Peng,C.-K., Simons, M. and Stanley,H.E. (1995) Long-Range Correlation Properties of Coding and Noncoding DNA Sequences: GenBank Analysis. Phys. Rev. E **51**, 5084-5091.

[21] Peng,C.K., Buldyrev,S., Goldberger,A., Havlin, S., Sciortino,F., Simons,M. and Stanley,H.E.(1992) Long-Range Correlations in Nucleotide Sequences. Nature **356**, 168-171.

[22] Li,W., Kaneko,K. (1992) Long-range correlations and partial 1/F-Alpha spectrum in a noncoding DNA sequence. Europhys. Lett. **17**, 655.

[23] Yang,H., Zhao,F., Zhuo,Y., *et al.* (2002) Analysis of DNA Chains By Means of Factorial Moments. Phys. Lett. A **292**,349-356.

[24] Provata, A., Almirantis,Y. (1997) Scaling properties of coding and non-coding DNA sequences. Physica A **247**, 482.

[25] Provata,A., Almirantis,Y. (2000) Fractal cantor patterns in the sequence structure of DNA, Fractals **8** ,15.

[26] Yang,H., Zhuo,Y., Wu,X. (1994) Investigation of thermal denaturation of DNA molecules

based upon non-equilibrium transport approach. J. Phys. A **27**, 6147-6156.

[27] Salerno,M. (1991) Discrete model for DNA-promoter dynamics. Phys. Rev. A **44**, 5292-5297.

[28] Lennholm,E., Homquist,M. (2003) Revisiting Salerno's sine-Gordon model of DNA: active regions and robustness. Physica D **177**, 233-241.

[29] Kalosakas,G., Rasmussen,K.O. and Bishop,A.R. (2004) Sequence-specific thermal fluctuations identify start sites for DNA transcription. Europhys. Lett. **68**, 127-133.

[30] Barral,J.P., Hasmy,A., Jimenez,J. and Marcano,A. (2000) Nonlinear modeling technique for the analysis of DNA chains. Phys. Rev. E **61**, 1812-1815.

[31] Xiao,Y., Huang,Y., Li,M., Xu,R. and Xiao,S. (2003) Nonlinear analysis of correlations in Alu repeat sequences in DNA. Phys. Rev. E **68**, 0619131-0619135.

[32] Farmer,J.D. and Sidorowich,J.J. (1987) Predicting chaotic time series. Phys. Rev. Lett. **59**, 845-848.

[33] Sugihara,G. and May,M. (1990) Nonlinear forecasting as a way of distinguishing chaos from measurement error in time series. Nature 344, 734-741.

[34] Rubin,D.M. (1992) Periodicity, randomness, and chaos in ripples and other spatial patterns. Chaos **2**, 525-535.

[35] Garcia,P., Jimenez,J. Marcano,A. and Moleiro,F. (1996) Local optimal metrics and nonlinear modeling of chaotic time series. Phys. Rev. Lett. **76**, 1449-1452.

[36] Suzuki,Y., Tsunoda,T., Sese,J., Taira,H., Mizushima-Sugano,J., Hata,H., Ota,T., Isogai,T., Tanaka,T., Nakamura,Y. et al. (2001) Identification and characterization of the potential promoter regions of 1031 kinds of human genes. Genome Res., **11**, 677-684.

[37] Matys,V., Fricke,E., Geffers,R., Gossling,E., Haubrock,M., Hehl,R., Hornischer,K., Karas,D., Kel,A.E., Kel-Margoulis,O.V. et al. (2003) TRANSFAC: transcriptional regulation, from patterns to profiles. Nucleic Acids Res., **31**, 374-378.

[38] Lenhard,B. and Wasserman,W.W. (2002) TFBS: computational framework for transcription factor binding site analysis. Bioinformatics, **18**, 1135-1136.

[39] http://repeatmasker.org/cgi-bin/WEBRepeatMasker. The options are all set default except the 'Repeat options' in the 'Advanced option', that the 'Only mask simple repeats and low-complexity DNA' is selected.

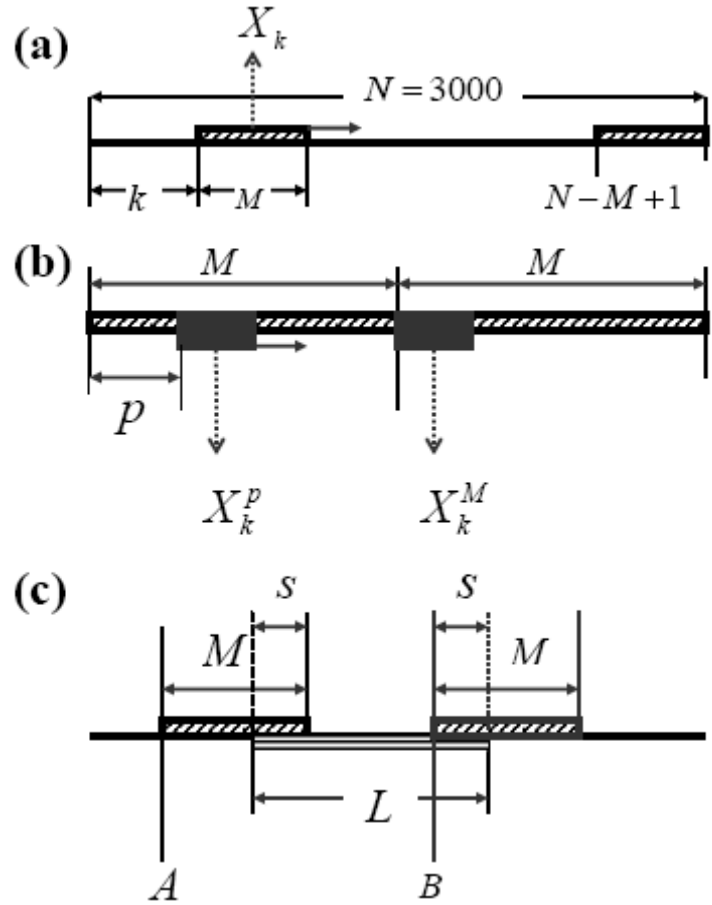

**Fig.1** Schematic representation of the nonlinear modeling (NM) method. (a) Sliding of a window with length $M$ along a PPR sequence can generate $N-M+1$ possible sub-chains. The statistical properties of each sub-chain can be employed to represent the local characteristics of the corresponding region. (b) For each sub-chain, connect its end with the starting of its replica. By this way we can obtain totally $M$ possible delay-register vectors with length $d$. From this set of delay-register-vectors we can obtain the statistical properties of the corresponding region. (c) Consider a segment with deterministic structure, the length of which is $L$. There is a delay interval that only when the overlap between the sliding window and the considered segment is larger than it can we find a distinguishable decrease of the overall average error. This interval is denoted as $s$.

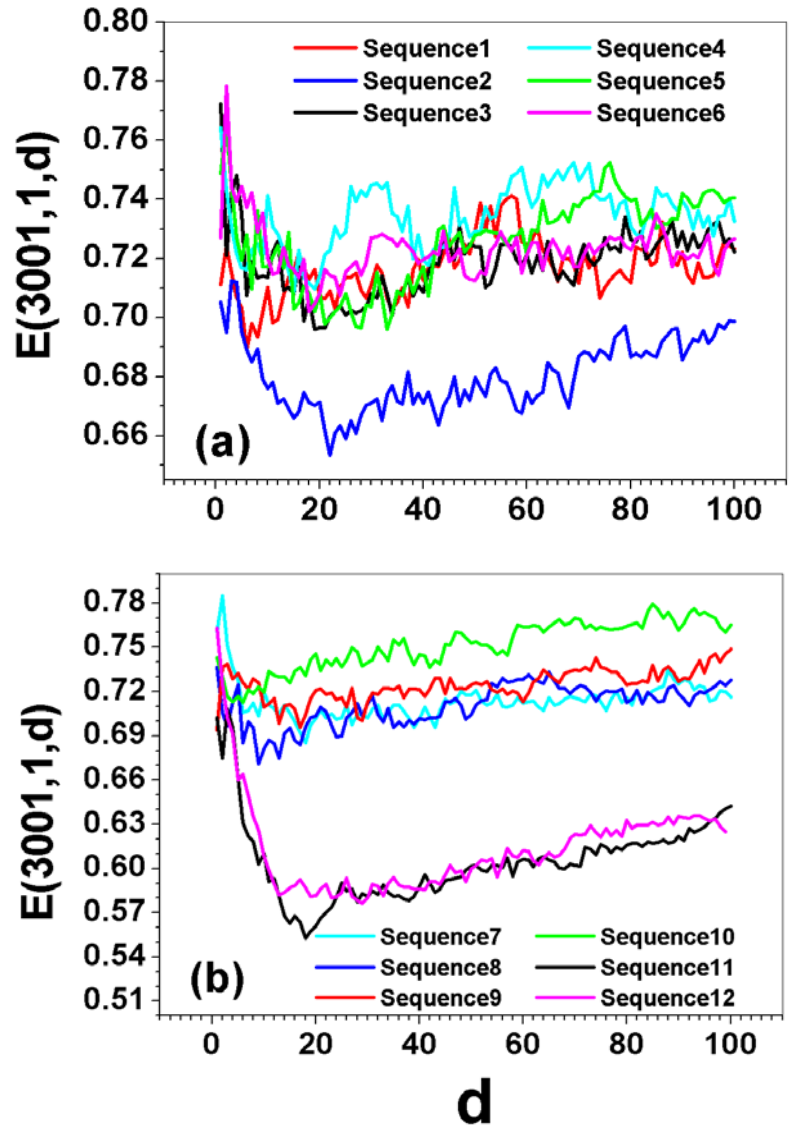

**Fig.2** (Color online) The overall mean error as a function of the embedding dimension $d$. The total length of a PPR is considered as a sub-chain, that is, $M = N$ and $k = 1$. The result for a PPR decreases rapidly with the increase of $d$ from 1 to a critical value $d_c$, where the overall error reaches the minimum value. From $d_c$ the overall error will increase slowly. The overall error in a dimension near $d_c$ can be employed as the representation of the deterministic structures in a PPR sequence. The results for the PPRs numbered 1 to 12 are presented.

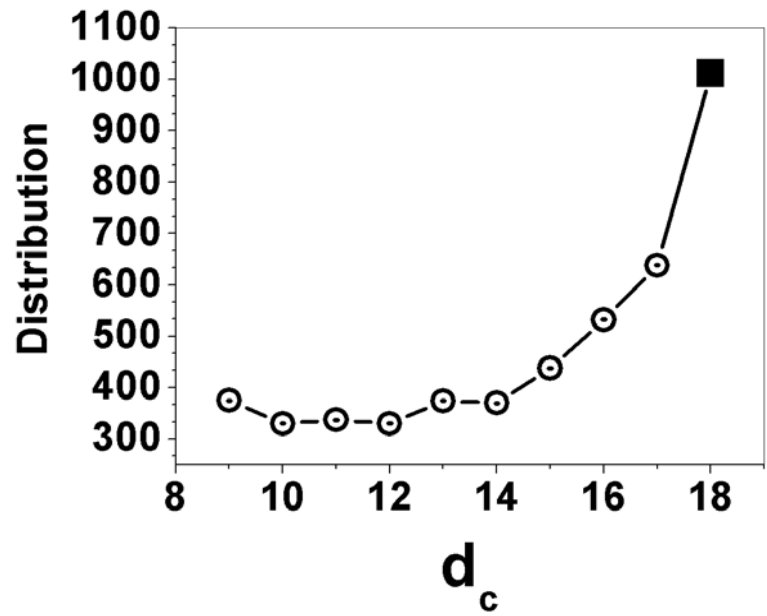

**Fig.3** The distribution of the critical value $d_c$ for the totally 4737 PPRs. The solid square denotes the number of the PPRs with $d_c \geq 18$ and $d_c < 9$.

**Table (1)** The Criterions to determine whether there is a deterministic structure in a DNA segment with a special set of occurrence probabilities for A, T, C and G. $(M, d, J) = (500, 15, 3000)$.

| p(A) | p(T) | p(C) | p(G) | $E_{uncorrelated}$ | $\dfrac{E_{criterion}(M,k,d)}{E_{uncorrelated}}$ | $\dfrac{\Delta(M,k,d)}{E_{uncorrelated}}$ |
|---|---|---|---|---|---|---|
| 0.10000 | 0.10000 | 0.10000 | 0.70000 | 0.4800 | 0.78717 | 0.21283 |
| 0.10000 | 0.10000 | 0.15000 | 0.65000 | 0.5350 | 0.83775 | 0.16225 |
| 0.10000 | 0.10000 | 0.20000 | 0.60000 | 0.5800 | 0.85522 | 0.14478 |
| 0.10000 | 0.10000 | 0.25000 | 0.55000 | 0.6150 | 0.92292 | 0.07708 |
| 0.10000 | 0.10000 | 0.30000 | 0.50000 | 0.6400 | 0.85802 | 0.14198 |
| 0.10000 | 0.10000 | 0.35000 | 0.45000 | 0.6550 | 0.88115 | 0.11885 |
| 0.10000 | 0.10000 | 0.40000 | 0.40000 | 0.6600 | 0.89876 | 0.10124 |
| 0.10000 | 0.15000 | 0.15000 | 0.60000 | 0.5850 | 0.86428 | 0.13572 |
| 0.10000 | 0.15000 | 0.20000 | 0.55000 | 0.6250 | 0.89563 | 0.10437 |
| 0.10000 | 0.15000 | 0.25000 | 0.50000 | 0.6550 | 0.86790 | 0.13210 |
| 0.10000 | 0.15000 | 0.30000 | 0.45000 | 0.6750 | 0.91469 | 0.08531 |
| 0.10000 | 0.15000 | 0.35000 | 0.40000 | 0.6850 | 0.89100 | 0.10900 |
| 0.10000 | 0.20000 | 0.20000 | 0.50000 | 0.6600 | 0.90238 | 0.09762 |
| 0.10000 | 0.20000 | 0.25000 | 0.45000 | 0.6850 | 0.90807 | 0.09193 |
| 0.10000 | 0.20000 | 0.30000 | 0.40000 | 0.7000 | 0.90702 | 0.09298 |
| 0.10000 | 0.20000 | 0.35000 | 0.35000 | 0.7050 | 0.89000 | 0.11000 |
| 0.10000 | 0.25000 | 0.25000 | 0.40000 | 0.7050 | 0.89746 | 0.10254 |
| 0.10000 | 0.25000 | 0.30000 | 0.35000 | 0.7150 | 0.90894 | 0.09106 |
| 0.10000 | 0.30000 | 0.30000 | 0.30000 | 0.7200 | 0.92045 | 0.07955 |
| 0.15000 | 0.15000 | 0.15000 | 0.55000 | 0.6300 | 0.87143 | 0.12857 |
| 0.15000 | 0.15000 | 0.20000 | 0.50000 | 0.6650 | 0.92282 | 0.07718 |
| 0.15000 | 0.15000 | 0.25000 | 0.45000 | 0.6900 | 0.91763 | 0.08237 |
| 0.15000 | 0.15000 | 0.30000 | 0.40000 | 0.7050 | 0.89669 | 0.10331 |
| 0.15000 | 0.15000 | 0.35000 | 0.35000 | 0.7100 | 0.92154 | 0.07846 |
| 0.15000 | 0.20000 | 0.20000 | 0.45000 | 0.6950 | 0.92770 | 0.07230 |
| 0.15000 | 0.20000 | 0.25000 | 0.40000 | 0.7150 | 0.89036 | 0.10964 |
| 0.15000 | 0.20000 | 0.30000 | 0.35000 | 0.7250 | 0.91623 | 0.08377 |
| 0.15000 | 0.25000 | 0.25000 | 0.35000 | 0.7300 | 0.91925 | 0.08075 |
| 0.15000 | 0.25000 | 0.30000 | 0.30000 | 0.7350 | 0.91117 | 0.08883 |
| 0.20000 | 0.20000 | 0.20000 | 0.40000 | 0.7200 | 0.91580 | 0.08420 |
| 0.20000 | 0.20000 | 0.25000 | 0.35000 | 0.7350 | 0.90870 | 0.09130 |
| 0.20000 | 0.20000 | 0.30000 | 0.30000 | 0.7400 | 0.89787 | 0.10213 |
| 0.20000 | 0.25000 | 0.25000 | 0.30000 | 0.7450 | 0.92362 | 0.07638 |
| 0.25000 | 0.25000 | 0.25000 | 0.25000 | 0.7500 | 0.92253 | 0.07747 |

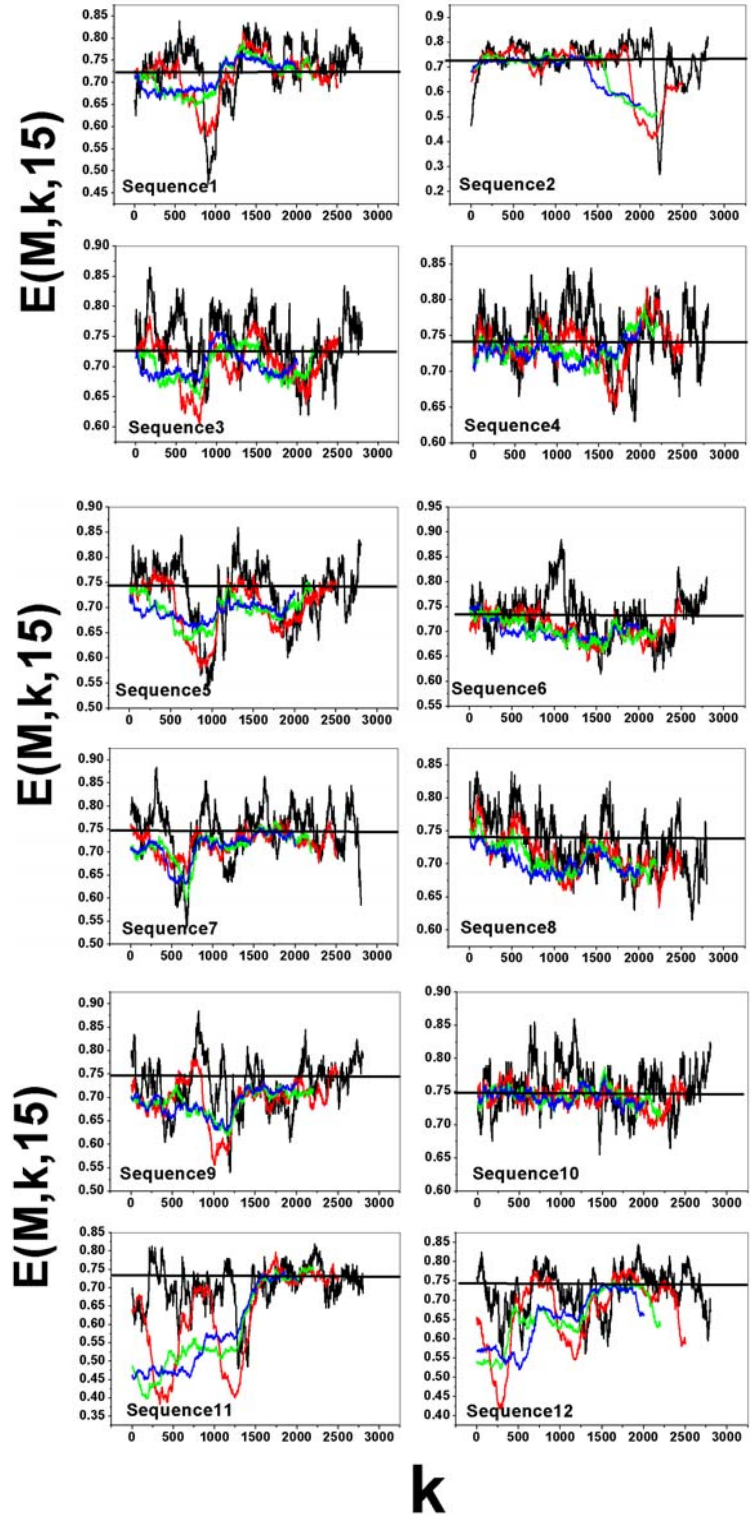

**Fig.4** (Color online) The overall error $E(M,k,15)$ versus the position $k$. Results for $M = 200, 500, 800, 1000$ are denoted with black, red, green and blue lines, respectively. The shuffling results of $E_{random}(3001,1,d)$ are presented with the straight level lines.

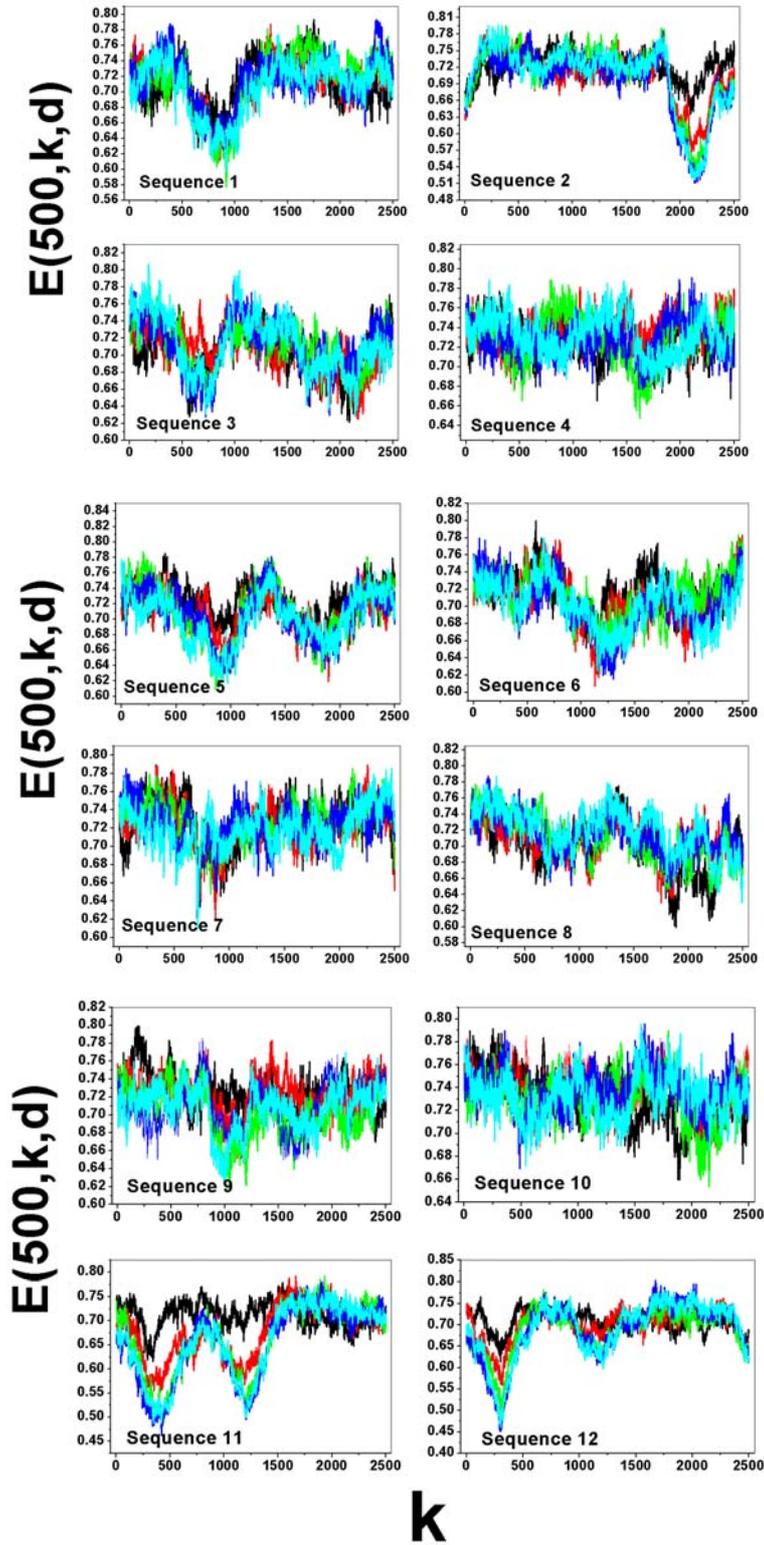

**Fig.5** (Color online) The overall error $E(500,k,d)$ versus the position $k$. Results for $d = 5, 10, 15, 20, 25$ are denoted with black, red, green, blue and cyan lines, respectively. The valleys are sensitive to the change of the dimension $d$.

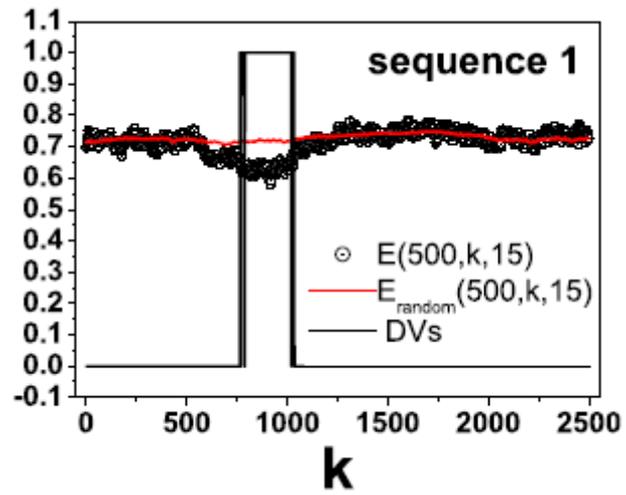

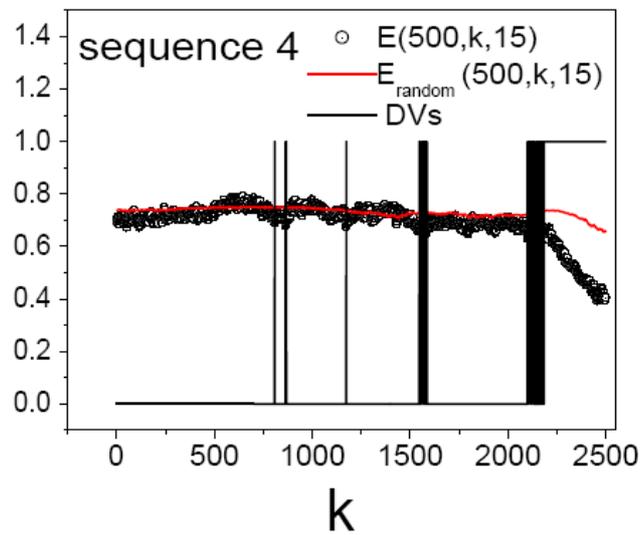

**Fig.6** (Color online)The valleys with deterministic structures are called DVs. The DVs can be identified by comparing the overall errors of the segments with the corresponding criterions. In the solid line the points whose values are 1 denote the segments having deterministic structures. A DV is a set of successive points in the solid line whose values are 1. Results for PPRs numbered 1 and 4 are presented as typical examples.

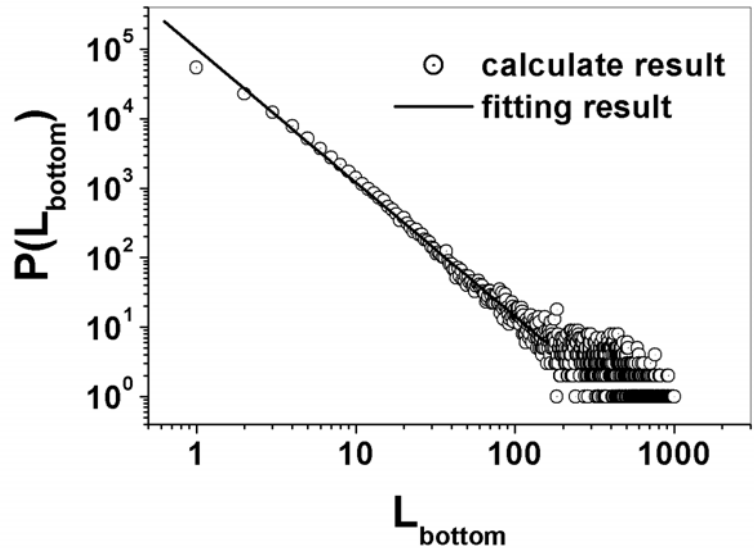

**Fig.7** The probability distribution function of the bottom lengths of the DVs（denoted with $L_{bottom}$）found in all the 4737 PPRS in the human genome. With the increase of length this probability decreases rapidly in a power-law way, i.e., $P_{DV}(L_{bottom}) \sim L_{bottom}^{-1.91}$. It is a nontrivial distribution function. It is reasonable to believe that the DVs should have some special biological functions.

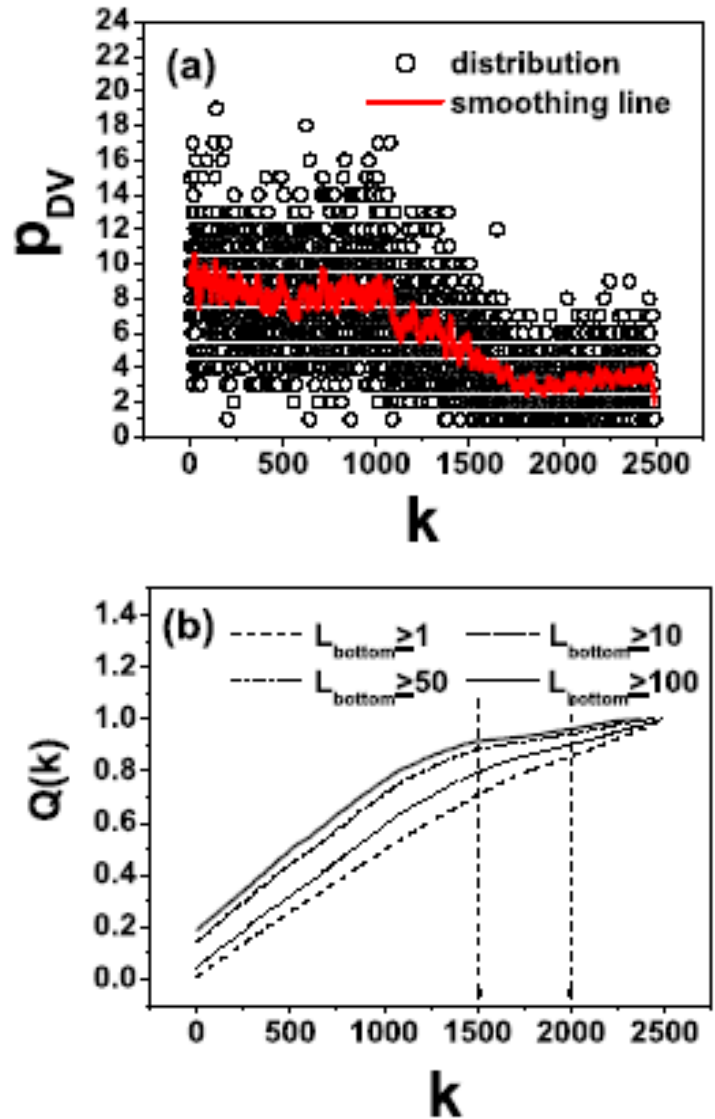

**Fig.8** (Color online) (a) The occurrence probability of the DVs at different positions. (b) The integrated probability of DVs at different positions. This occurrence probability is obtained by using all the DVs found in the 4737 PPRS in the human genome. There is a special region around the transcriptional start site, i.e., [1500,2000], where the integrated probability increases with a significantly slow speed. Consequently, the occurrence probabilities in this region should be small significantly comparing with that outside of this region. The DVs with long bottom lengths prefer much more to occur outside of this special region. Comparison with the occurrence frequency of CpG islands tells us that the DVs should have special biological functions rather than CpG islands.

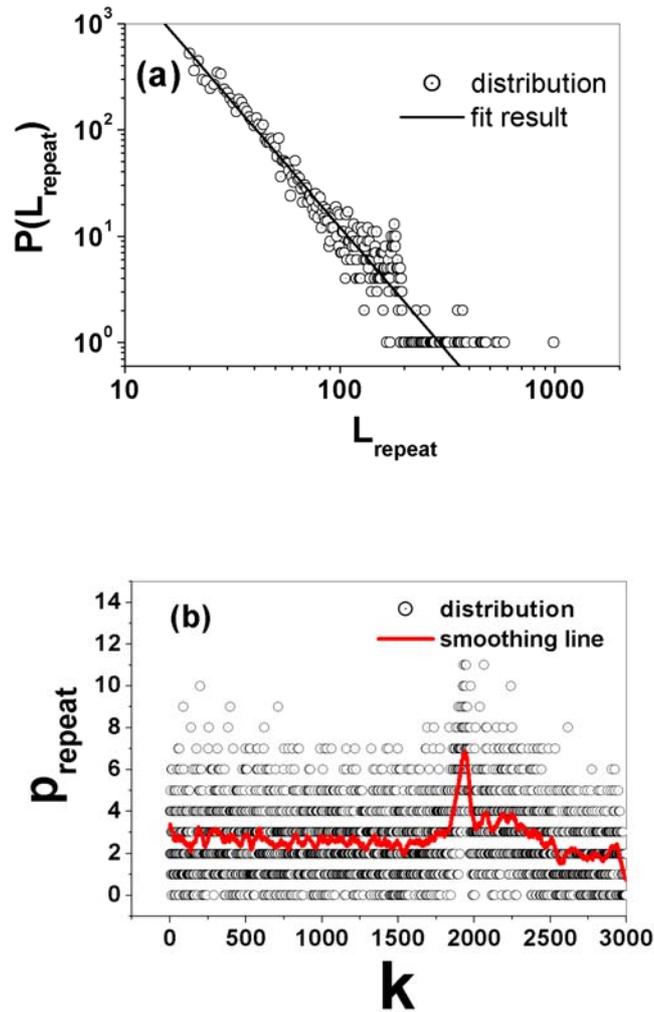

**Fig.9** (Color online) (a) The length distribution of the repetitive or low-complexity sequences. It obeys a power-law, i.e., $P(L_{repeat}) \propto L_{repeat}^{-2.35}$. (b) The occurrence probability at different positions $p_{repeat}(k)$. It distributes homogenously in all the positions except the special range $[1850,1980]$, where the proximal promoter regions prefer to occur and a significant sharp peak is shaped.